# Change of the resonant electron orbit from trapped orbit to passing orbit in fast wave current drive

Yao Kun, Zhao Yanping

Institute of Plasma Physics, Chinese Academy of Sciences, Hefei, 230031, China

**Abstract** In fast wave current drive, the resonant electron is accelerated by fast wave in the direction parallel to the static magnetic field, and the parallel velocity will be increased. The trajectories of the trapped resonant electrons are calculated with a computer code in which fast wave-induced diffusion in velocity space is accounted for by a quasi-linear operator. The simulation results show that the orbit of trapped resonant electron will change from a trapped orbit to a passing orbit in some cases. We obtain the transition conditions, and if they are satisfied the trapped orbit will become a passing orbit. The transition from trapped orbit to passing orbit implies that the effect of trapped electrons on current drive will be reduced and the current drive efficiency will be improved.

**Keywords:** fast wave current drive, resonant electrons, orbit.

**PACS:** 52.20.Dq, 52.55.Fa, 52.55.Wq

## 1. Introduction

Non-uniformity of magnetic field forms the magnetic wells in tokamak. Electrons with a small enough parallel velocity relative to their perpendicular velocity, will bounce off the magnetic wells in their motion along the field lines. These electrons are called trapped electrons. In Radio-Frequency (RF) current drive, trapped electrons may degrade current drive efficiency because they absorb wave energy without producing the net toroidal current in a time averaged sense. The effect of trapped electrons on current drive efficiency has been studied [1-6]. In kinetic theory, the bounce averaging procedure allows us to take into account the effect of trapped electrons on the RF driven current [7-9].

In tokamak, particle orbit can transit from one orbit class to another. In ion cyclotron heating, as the ion passes through the cyclotron layer and undergoes random "kicks" [10], there is a continuous flow of particles from the passing region to the trapped region and vice versa. This leads to the ion radial diffusion [11]. In electron cyclotron resonance heating (ECRH), the wave pushes electrons in perpendicular direction and the perpendicular velocity increases, this leads that the electron orbit changes from a passing orbit to a trapped orbit.

In fast wave current drive, the force exerted on electron by wave is in the parallel direction. If the force is large enough that the trapped electron can overcome the magnetic well, the resonant trapped electron can become a passing electron. In this paper, we will show that the resonant trapped electron can become a passing electron when the diffusion coefficient is large enough.

## 2. Simulation model

### 2.1 Basic assumptions

In order to prevent the important physical phenomena masked by possible computational and conceptual intricacies, we construct the simplest prototype model that describes the resonant electron motion in fast wave current drive. In calculation, we assume the fast wave amplitude to be fixed in space and time. Pitch-angle scattering and slowing-down collisions are neglected. The static electric field is zero.

### 2.2 Representation of equilibrium magnetic field

The contravariant and covariant expressions of the equilibrium fields in an axisymmetric torus are

$$B = \nabla \zeta \times \nabla \psi + q \nabla \psi \times \nabla \theta \tag{1}$$

$$B = g(\psi)\nabla \zeta + I(\psi)\nabla \theta + \delta(\psi,\theta)\nabla \psi \tag{2}$$

where ($\psi$, $\theta$, $\zeta$) are the magnetic flux coordinates, with $\psi$ the poloidal magnetic flux, $\theta$ the poloidal angle, and $\zeta$ the toroidal angle.

In plasma equilibrium, the Grad-Shafranov equation can be written as follows:

$$R\frac{\partial}{\partial R}\left(\frac{1}{R}\frac{\partial \psi}{\partial R}\right) + \frac{\partial^2 \psi}{\partial Z^2} = -\mu_0 R J_\phi \tag{3}$$

From (3), we can obtain the poloidal magnetic flux [12].

### 2.3 Hamiltonian guiding center equations

Electron trajectories are calculated by solving the equations [13]:

$$\frac{dP_\theta}{dt} = -\frac{\partial H}{\partial \theta} \tag{4}$$

$$\frac{d\theta}{dt} = \frac{\partial H}{\partial P_\theta} \tag{5}$$

$$\frac{dP_\alpha}{dt} = -\frac{\partial H}{\partial \alpha_c} \tag{6}$$

$$\frac{d\alpha_c}{dt} = \frac{\partial H}{\partial P_\alpha} \tag{7}$$

where

$$\frac{\partial Q(\psi,\theta)}{\partial \psi} = \frac{\partial \delta(\psi,\theta)}{\partial \theta} + \frac{dq}{d\psi} \tag{8}$$

$$\alpha_c = -\zeta + q\theta - \int_{P_\alpha}^{\psi} \left[\delta(\psi^*,\theta) + \frac{dq}{d\psi}\theta\right] d\psi^* \tag{9}$$

$$h(\psi) = q(\psi)g(\psi) + I(\psi) \tag{10}$$

$$P_\theta = \rho_{//}h - \rho_{//}gQ + \int_{P_\alpha}^{\psi} Q(\psi^*,\theta)d\psi^* \tag{11}$$

$$P_\alpha = \psi - \rho_{//}g \tag{12}$$

$$H = \frac{1}{2}\rho_{//}^2 B^2 + \mu B \tag{13}$$

$$\rho_{//} = \frac{mv_{//}}{Bq} \tag{14}$$

$$\mu = \frac{1}{2}m\frac{v_\perp^2}{B} \tag{15}$$

## 2.4 Diffusion operator

The effect of the fast wave on the parallel velocity of the resonant electron is modeled with a Monte Carlo operator acting over time steps. The operator is given by [14]

$$\Delta\lambda = -2(\lambda - \lambda^3)D\Delta t \pm \sqrt{(1-\lambda^2)^2 D\Delta t} \tag{16}$$

where

$$\lambda = \frac{v_{//} - (v_{//1} + v_{//2})/2}{(v_{//2} - v_{//1})/2} \tag{17}$$

$$D = \begin{cases} D_0, v_{//1} < v < v_{//2} \\ 0, otherwise \end{cases} \tag{18}$$

$$D_0 = \frac{B_{//}^2}{B_0^2} k_{//} (2v_t^2 - v_\perp^2)^2 \pi \frac{4}{(v_{//2} - v_{//1})^3} \tag{19}$$

B// is the fast wave field parallel to the magnetic field, and $B_0$ is the equilibrium magnetic field on the axis. $v_t$ is the electron thermal velocity. $k_{//}$ is the mean wave number of the wave spectrum. The symbol $\pm$ means the sign is to be chosen randomly, but with equal probability.

## 3. Simulation results and discussion

### 3.1 Simulation setup

The wave amplitude and electron temperature profiles are of the form $(1-\rho^2/a^2)^{1/2}$. $\rho$ is defined as the half width of a surface at the equator and a is the minor radius. The plasma and fast wave parameters are given in **table.1**. The equilibrium magnetic surfaces used in this paper are shown in **figure.1**. We simulate a typical test electron motion on a 129×65 ($\psi$, $\theta$) grids. The initial electron energy is chosen to be 10 kev, the initial position is (1.9799m, 0.0000m), and the initial pitch angle is 72 degree. This electron trajectory is shown in **figure 2**. It is a deeply trapped electron, and the finite width of the trapped orbit is very small because the width is proportion to particle mass.

| Plasma parameters | Wave parameters |
|---|---|
| $R_0$=1.7 m | $v_{//1}=10^7$ m/s |
| a=0.4 m | $v_{//2}=10^8$ m/s |
| $\kappa$=1.88 | $k_{//}=3$ m$^{-1}$ |
| $\delta$=0.75 | |
| $I_p$=1 MA | |
| $B_0$=3.5T | |
| $T_{e0}$=5 kev | |

**TABLE. 1 Range of parameters for simulations**

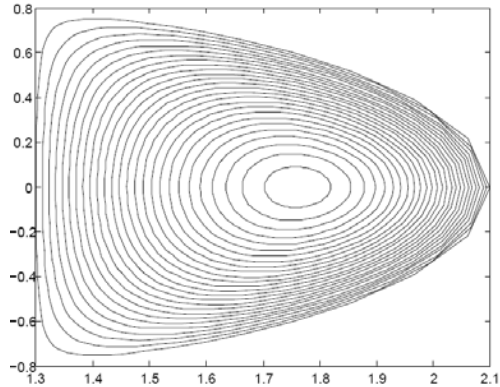

**Figure.1 Equilibrium magnetic surfaces**

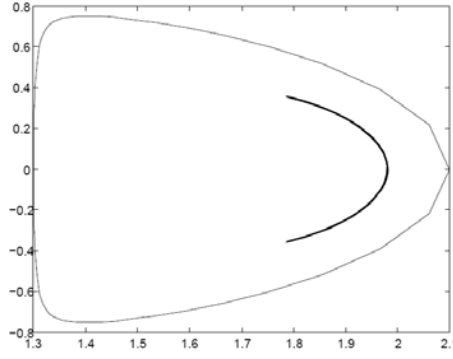

**Figure.2 Electron trajectory without fast wave**

**3.2  Results and discussions**

The electron trajectories are calculated in different wave amplitudes. The electron trajectories are depicted for three cases in **figure.3**. For the first two cases (**figure.3 (a) and (b)**), the trapped resonant electron becomes a passing electron because the fast wave amplitude is large enough. The difference between the first two cases is that the resonant electron orbit in the first case changes from trapped orbit to passing orbit faster than it does in the second case. For the first case, the electron becomes a passing electron after 1000 time steps. For the second case, the electron becomes a passing electron after 5000 time steps. For the third case (**figure.3 (c)**), the fast wave amplitude is so small that the trapped electron can't overcome the magnetic well after 10000 time steps.

We write the force equation for resonant electrons as[15]

$$m_e \frac{dv_{//}}{dt} = \langle F_w \rangle + \langle F_g \rangle \tag{20}$$

$$F_w = \frac{e}{2\Omega_e}(2v_t^2 - v_\perp^2)\frac{\partial B_{//}}{\partial z} \tag{21}$$

$$F_g = -\mu \nabla_{//} B \tag{22}$$

$F_w$ is the force exerted on electron by fast wave, and $F_g$ is the force arising from the gradient of the static magnetic field. < > denotes time average. In a bounce period $T_b$, $\langle F_g \rangle$ is about zero. After the time $t=nT_b$, the increment of parallel velocity can be written as

$$\Delta v_{//} = \frac{\langle F_w \rangle}{m_e} t \tag{23}$$

The trapped electron satisfies

$$\frac{B_{min}}{B_{max}} < \left(\frac{v_{0\perp}}{v_0}\right)^2 \qquad (24)$$

As the increase of the parallel velocity, the trapped resonant can become a passing electron and satisfy

$$\frac{B_{min}}{B_{max}} \geq \frac{v_{0\perp}^2}{v_0^2 + (\Delta v_{//})^2} \qquad (25)$$

From (25), we can obtain

$$t \geq \sqrt{\frac{v_{0\perp}^2 - v_0^2 B_{min}/B_{max}}{B_{min}\langle F_w\rangle^2/(B_{max} m_e^2)}} \qquad (26)$$

Obviously, as long as the time is long enough, the trapped resonant electron can become a passing electron no matter how small the fast wave amplitude is. In the third case, if we increase the time steps for simulation, we will find that the trapped electron becomes a passing electron after 100000 time steps. **Figure.3 (d)** shows this result. But if the perpendicular velocity is very large relative to $v_{//2}$ and it satisfies

$$\frac{B_{min}}{B_{max}} < \frac{v_{0\perp}^2}{v_{0\perp}^2 + v_{//2}^2} \qquad (27)$$

the trapped resonant electron can't become a passing electron. The resonant electron is accelerated by fast wave, its velocity can be beyond the resonant velocity region, and it is no longer accelerated by fast wave.

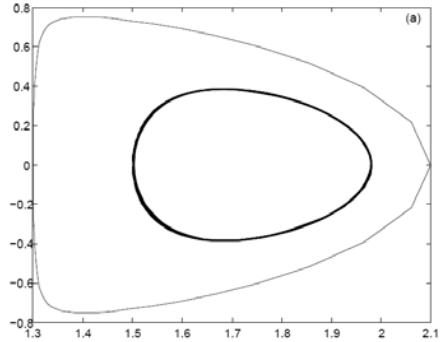

(a) $B_{//}/B_0 = 0.01$

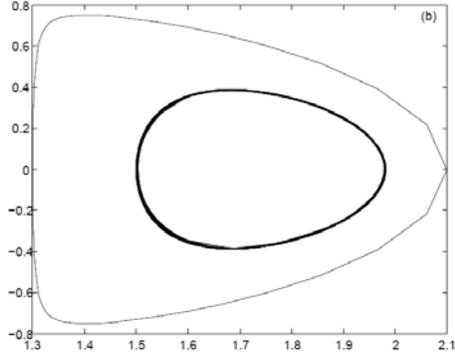

(b) $B_{//}/B_0 = 0.001$

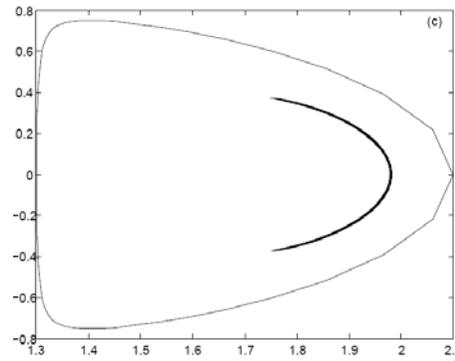

(c) $B_{//}/B_0 = 0.0001$

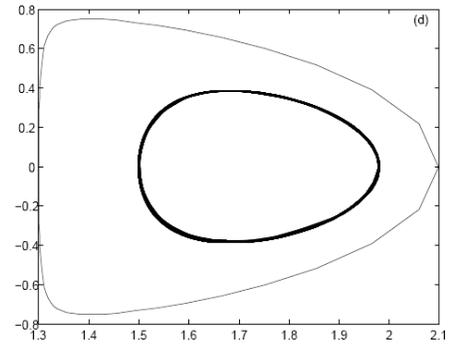

(d) $B_{//}/B_0 = 0.0001$

**Figure.3 Electron trajectories with different fast wave amplitude**

## 4. Summary and conclusions

In fast wave current drive, Landau damping and transit time magnetic pumping induce two parallel forces respectively, and these two forces are opposite. When $v_\perp^2 = 2v_t^2$, there is an exact cancellation of these two forces and the diffusion coefficient is zero. So for the trapped resonant electron with $v_\perp^2 = 2v_t^2$, it can't become a passing electron.

In the outer part of magnetic surfaces, the inverse aspect ratio is not small and a large fraction of electrons are trapped. These trapped electrons will degrade the fast wave current drive efficiency. But the simulation results show that a fraction of the trapped resonant electrons can become passing electrons after a period of time in fast wave current drive. In actual situations the current drive time is finite. If the fast wave amplitude is too small, the trapped resonant electron will not become a passing electron within the current drive time. In order to reduce the effect of trapped electrons on current drive, we need prolong the current drive time, enhance the power of fast wave and heighten the upper limit of resonant velocity region.

**Acknowledgements**

Thanks for Xiao's help in computer code and helpful discussions.